\documentclass{aa}  

\usepackage{graphicx}
\usepackage{subcaption}
\usepackage[colorlinks=true,linkcolor=blue,citecolor=blue]{hyperref}

\usepackage{txfonts}

\begin{document}

   \title{Machine learning for the early classification of \\ broad-lined Ic supernovae }
    \titlerunning{Machine learning for early Ic-BL classification}
    \authorrunning{L. Cotter et al.}

   \author{L. Cotter,
          \inst{1}
          A. Martin-Carrillo \inst{1},
        J. Fisher \inst{1},
        G. Finneran  \inst{1},
        G. Corcoran  \inst{1},
        J. Lebron  \inst{2}
          }

   \institute{School of Physics and Centre for Space Research, University College Dublin, Belfield, Dublin 4, Ireland\\
              \email{laura.cotter@ucdconnect.ie}
         \and
             Computing Faculty, Griffith College, Dublin 6, Ireland
             }

   \date{Received 19 December 2025 / Accepted 18 March 2026}

  \abstract
{
  Science is currently at an age where there is more data than we know how to deal with. Machine learning (ML) is an emerging tool that is useful for drawing valuable science out of incomprehensibly large datasets and identifying complex trends in data that may otherwise be overlooked. Moreover, ML can potentially enhance the quality and quantity of scientific data as they are collected. This paper explores how a new ML method can improve the rate of classification of rare broad-lined Ic (Ic-BL) supernovae (SNe). We introduce new parameters called magnitude rates to train ML models to identify SNe Ic-BL in large datasets and apply this same methodology to a population of SN Ia to test if our ML approach is reproducible. The information we required to train each ML model included three magnitudes, three time differences, two magnitude rates, and the second derivative of these rates using the first three available photometric data points in a single filter.  Our initial investigations showed that the random forest algorithm provides a strong foundation for the early classifications SNe Ic-BL and SNe Ia. Testing this model again on an unseen dataset showed that the model can identify upward of  13.6\% of the total true SN Ic-BL population, significantly improving on current methods. By implementing a dedicated observation campaign using this model, the number of SN Ic-BL classified and the quality of early-time data collected each year will see considerable growth in the near future.}

   \maketitle

\section{Introduction}

Supernovae (SNe) are among the most energetic and extraordinary phenomena in our Universe. They mark the end of a massive star's life, exploding with an enormous burst of energy while also ejecting matter and radiation into the interstellar medium at high velocities. There are two main groups of SNe, Type I and Type II, and each is further subdivided into sub-classes with its own characteristics \citep{Mink}.

The diversity of SNe events is believed to primarily be governed by the initial mass and the mass loss rates of their progenitor stars, along with their characteristic spectral features \citep{Filip,SN_form}. This paper focuses on a rare subclass of Type I SNe known as broad-lined Ic SNe (SNe Ic-BL). These transients are fully stripped core-collapse SNe whose spectra exhibit no hydrogen or helium lines and display broad-line features, which are indicative of ultra-high expansion velocities. {Some SNe Ic-BL have known associations with gamma-ray bursts \citep[GRBs; e.g.][]{galama1998, SNGRB, cano2017}. Additionally, their multi-wavelength monitoring from early times has shown great potential for their use as probes to search for off-axis GRB jets \citep[e.g.][]{izzo2020, schroeder2025}, making these SNe a topic of great interest amongst astronomers.

Notably, SNe Ic-BL are very rare, with on average only $\sim$\,20 spectroscopically classified per year. Based on SN population studies \citep[e.g.][]{pessi25}, it is believed that many more SNe Ic-BL are being detected but are not being spectroscopically confirmed, as only a few thousand, out of tens of thousands of possible SNe transients detected each year, are officially classified. While the classification of SNe is not entirely unbiased, with SNe Ia favoured due to their brightness compared with other SNe, it can be assumed that their classification is almost random. Accordingly, the classification rates at a first order of magnitude between the different SN classes can be compared to their true population rates and can be used to predict the approximated expected percentage of SNe that were detected in a year but not classified. According to the Transient Name Server (TNS),\footnote{\url{https://www.wis-tns.org}} SNe Ic-BL typically only make up 0.8$\%$ of the total classified sample. As a result, each year, it is expected that over 150 SNe Ic-BL could be detected but missed due to not making it to the classification process.

Among the limited number of SNe Ic-BL detected each year, even fewer have reliable photometric data, if any, to build an acceptable light curve. When considering the quality of spectroscopic and photometric data, the number of well-documented SNe Ic-BL drops to approximately ten or fewer annually. If we also consider the number of SNe Ic-BL with adequate early rising time light curve data, this number drops again to around five. The limited availability of early high-quality data coupled with the limitations of current SN classification methods significantly hampers the ability to conduct the science required to address the many unresolved questions surrounding these transients. Changes to the classification processes for SNe are crucial to prevent the oversight of valuable early-time scientific data \citep{Mod19}.

Machine learning (ML) algorithms such as those employed by the \texttt{ALeRCE} broker assign classifications to SNe detections in large datasets.\footnote{\url{https://ALeRCE.science}} With SNe Ia being the largest population of any SN class, they are the most obvious choice for training sets, and as a result, SNe Ia may be favoured in the classification process \citep{MLIa}. Type Ib, Ic, and II SNe samples are also used in developing and training ML algorithms \citep{ML1,ML2}. The SNe Ic-BL, however, have never been looked at in isolation due to their small population.

Improving and training ML models with SN Ic-BL datasets is crucial to increasing the number and timeliness of SNe Ic-BL classified each year. Current ML algorithms are classifying SNe too late, resulting in a loss of valuable early-time data. SNe Ic-BL light curves rise rapidly compared to other SN classes, boasting expansion velocities as high as 0.1 times the speed of light during this phase \citep{modjaz,Lc_temp, gabriel}. This rapid rise makes it even more difficult to gather early-time data for these transients, leading current methods to fail in producing prompt classifications.

This classification delay also hampers the study of the SN-GRB connection in Ic-BL SNe. In cases where an SN is associated with a GRB, the initial stage of the explosion involves the bipolar jet of the GRB breaking through the surface of the progenitor star, which forms a hot cocoon around the jet \citep{Izzo}. At later times, the $^{56}$Ni produced within the progenitor star powers the radioactive heating of the ejecta, resulting in the observed SN emission \citep{Arnett}.

Gamma-ray bursts occur on short timescales that range from several milliseconds to several hours \citep[e.g.][]{woo2}. %This means that the initial GRB part of the GRB-SN interaction also occurs very quickly. 
The GRB jets are narrow (with opening angles $\sim 10-20^\circ$) \citep[e.g.][]{wang2020}, which can lead to the prompt $\gamma$-ray emission being missed when observed at viewing angles greater than the jet opening angle. The presence of these GRB jets can still be inferred by their non-thermal afterglow emission as the jet decelerates and spreads \citep[e.g.][]{granot2002, kumar2003, piran2004}. In these cases, the SN emission can obscure or rival these faint afterglows in certain wavelengths, especially in the optical band. To effectively capture potential GRB contributions, the focus must be on observing the early stages of the SN before it reaches its peak brightness \citep[e.g.][]{barnes, izzo2020}. Early classifications are desperately required to achieve this, which current methods are struggling to produce \citep{ztf22, schroeder2025}.

The need for prompt classifications of SNe Ic-BL to observe a possible SN-GRB connection is the primary motivation for this paper. In Sect.~\ref{sec:MLZTF} we show how SNe with the tag SN Ibc obtained from the \texttt{ALeRCE} broker are compared to their actual classifications from TNS \citep{ALeRCE_api} and the limitations that this first ML layer imposes. In Sect.~\ref{sec:SNparam} we propose a new parameter space in which the early rising rates of the light curves between the different classes of SNe provide key information to segregate Ic-BL from the rest of the SN population. In Sect.~\ref{sec:sample} we describe the Ic-BL selection used for the ML training described in Sect.~\ref{sec:ML}. The results and conclusions of this ML approach, which are useful for identifying newly discovered Ic-BL SNe candidates, are included in Sec.~\ref{sec:results} and Sec.~\ref{sec:conclusions}.

\section{Motivation and tests on the \texttt{ALeRCE} classifier}
\label{sec:MLZTF}
\subsection{\texttt{ALeRCE} data processing}

The Automatic Learning for the Rapid Classification of Events (\texttt{ALeRCE}) pipeline ingests transient data from the Zwicky Transient Facility (ZTF) alert stream, which uses a combination of different ML techniques to classify different types of transients \citep{ALeRCE_api}. The \texttt{ALeRCE} pipeline processes data firstly by using the information in the ZTF alerts about objects surrounding the transient, and this information is then cross-matched with other catalogues. As new transients are detected, ZTF uses a stamp-based classification system, which comprises a rotationally invariant convolutional neural network. One of five labels -- SN, active galactic nuclei (AGN), variable star (VS), asteroid, or bogus -- is assigned to the new transient \citep{stamp}. The ZTF alert data then undergo pre-processing, where the magnitudes of the object are corrected. Light curves of detected transients are examined for features in order to narrow the classification to one of 15 classes using various classification models. The SNe classes that are included are SN Ia, SN Ibc, SN II, and super luminous (SL) SNe \citep{san}. The classifications of transients from their light curves are made with a minimum of six detections obtained in either the \textit{g} or the \textit{r} band.

\subsection{Reliability of the \texttt{ALeRCE} SN Ibc classification}
From \texttt{ALeRCE}, all of the transients detected between 2018 and 2022 with the classification of SN Ibc were queried and matched to TNS transients using the RA and DEC of the objects. The reliability of the \texttt{ALeRCE} classifications was tested by comparing these classifications to their official SNe classifications. Initial investigations revealed 77.8\% of the transients were incorrectly given the label SN Ibc by the \texttt{ALeRCE} system. Each transient in the \texttt{ALeRCE} system has a list of probabilities of that transient being a given class. To obtain a more representative sample of the contamination in the \texttt{ALeRCE} SN Ibc classifications, we created a smaller dataset that contained all transients with a probability greater than 0.45 of being an SN Ibc. The contamination fell to 57.9\% incorrectly classified SN Ibc transients when considering this smaller dataset.

\subsection{Monitoring of the \texttt{ALeRCE} machine learning classification}

To investigate the ML approach used by \texttt{ALeRCE}, we developed a Python monitoring script to track transients as alerts entered the \texttt{ALeRCE} database. The script initially collected the 200 most recent SN Ibc transients and performed daily checks for newly labelled events, monitoring how their subtype classifications evolved as additional data became available. For each transient, we recorded information on detections, classifications, class probabilities, and the number of data points in the \textit{g} and \textit{r} bands at a daily cadence. In the four months that the script was run, many transients oscillated between classifications, with changes occurring even months after the first detection. Such unstable classifications present a significant challenge for identifying specific SN classes at early times. Additionally, we noticed a lack of high-probability SN Ibc classifications, with ML probabilities rarely exceeding $\sim 0.6$, even for confirmed events. These limitations highlight the need for a more targeted and robust ML classification approach.

\section{Exploration of a new parameter space for machine learning supernova classification}
\label{sec:SNparam}

\subsection{Approaching new machine learning methods}
Most of the ML approaches described in the literature are aimed at SN Ia, and in general, they rely on the complete set of photometric information gathered from a detailed SN monitoring campaign \citep[e.g.][]{Lochner2016}. Using a set of different statistical features, \cite{ml_old} attempted to use ML to classify SNe Ia and core-collapse SNe first at early epochs (focusing on the initial part of the light curve) and then using the entire SN light curve. For early epoch classifications, the results revealed the average slope feature as one of the most important features in their ML approach.\footnote{\cite{ml_old} was withdrawn before all referee comments could be fully addressed, and thus the findings should be interpreted with appropriate caution.} In this study, we use a similar feature called a magnitude rate. This magnitude rate is the difference in magnitude between two consecutive data points on a light curve per unit of time, which is calculated as

\begin{equation}
    \text{Magnitude rate} = \frac{\text{mag}_2 -\text{mag}_1 }{\text{time}_2 - \text{time}_1}.
\end{equation}

Additionally, the second derivative of the rates can be calculated using two magnitude rates and is calculated as follows:

\begin{equation}
    \text{Second derivative} = \frac{\text{Mag Rate}_2 -\text{Mag Rate}_1 }{\text{time}_3 - \text{time}_1}.
\end{equation}

These features reveal information on the rate at which the SN light curve rises and could be the key to improving early epoch classifications, as it only requires three photometric points of a transient as input.

\subsection{Light curve fitting}
\label{sec:maths} % used for referring to this section from elsewhere

The potential value of the magnitude rates was investigated by fitting full light curves to 402 SN Ibc \texttt{ALeRCE} classified transients with a probability of being an SN Ibc greater than 0.45. The light curve equation is described as

\begin{equation}
\label{light}
    m(t) = \frac{y_0 + m(t-t_0) + g_0\text{exp}(-(t-t_0)^2/2\sigma^2) }{1 - \text{exp} ((\tau -t)/\theta)}.
\end{equation}

Here $y_0$ is the intercept of the linear decay of the tail of the light curve, which is described by a slope (\textit{m}). The second term in the numerator describes the normalised Gaussian peak with amplitude ($g_0$), phase ($t_0$), and width ($\sigma$). 
The denominator describes the exponential rising observed at the beginning of the light curve, with $\theta$ being the characteristic time and $\tau$ being the phase zero point \citep{Carnegie}.

A minimum of three light curve data points was required for the SNe light curve fits, which further reduced the sample to 339 transients. The transient light curves were fit in their entirety rather than just the rising, ensuring that the shape of the rise was consistent with the rest of the light curve, which allowed for a true representation of an SN light curve.

\subsection{Comparing magnitude rates of SN classes}

The fitted light curve parameters allow the magnitude rates to be calculated at any time. The magnitude rate per day was calculated for each SN, and the SNe were then grouped into their respective classes. The curve shown in figure~\ref{comp} corresponds to the median of the magnitude rates at each pre-peak time interval for each SN class, fitted to an exponential function described as

\begin{equation}\label{exp}
    f(x) = a +b\text{exp}(-cx).
\end{equation}

\vspace{3pt }

\begin{figure}[h!]
    \centering
    \includegraphics[width=0.34\textwidth]{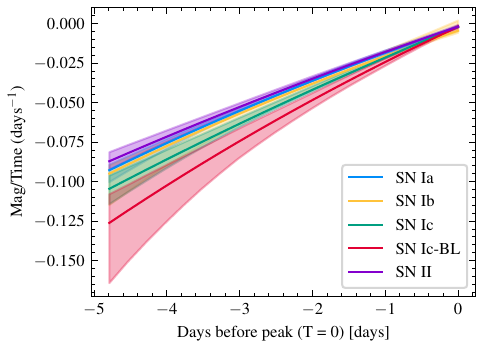}
    \caption{Fits to the median magnitude rates for different SN classes, evaluated 5 days before the peak of the light curve. Uncertainties correspond to the 90\% credible interval derived from 1000 posterior samples.}
    \label{comp}
\end{figure}

As seen in figure~\ref{comp}, SN Ia, Ib, and II all have similar magnitude rate curves, making it difficult to distinguish between them. The SN Ic-BL curve, however,  clearly deviates from the other curves at very early times, suggesting that for SNe with very early detection points, the magnitude rate can be used as a tool to separate SNe Ic-BL from the rest of the population.

\section{Supernova sample}
\label{sec:sample}

We built the largest sample of SNe Ic-BL by collecting all public data in different optical bands, resulting in 205 SN Ic-BL datasets. These datasets contain photometric points, redshifts, positions, discovery dates, and discovery groups.

Following a similar approach to the sample built for Sect.~\ref{sec:maths}, we queried all SN transients in \texttt{ALeRCE} from 2018 to 2024. These SNe were matched to classified TNS transients using RA and DEC. A key condition was that each transient had to have good risings ahead of the SN peak and at least three points (including the SN peak) in the early rising. In total, 4435 individual SN were included in this sample, with 7220 multi-band light curves. Of these transients, 136 individual SNe Ic-BL transients met our requirements, resulting in 265 multi-band SN Ic-BL light curves.

\section{Methodology}
\label{sec:ML}
\subsection{Using scikit-learn}

The Python module \texttt{scikit-learn} was employed to train the ML algorithm on our training sample. The basis of this module includes other well-known modules such as \texttt{NumPy} \citep{harris2020array}, \texttt{SciPy} \citep{2020SciPy-NMeth}, and \texttt{matplotlib} \citep{Hunter:2007}, making the ML technique accessible and straightforward to use by those who have limited working knowledge of ML but are fluent in Python. As an open-source ML library, \texttt{scikit-learn} provides a wide range of built-in ML algorithms and other supporting tools that enable users to efficiently select, evaluate, and fit models \citep{scikit-learn}.

The module enables a binary classification task to be performed on the sample of 339 SN Ibc-classified transients from \texttt{ALeRCE} with a probability equal to or greater than 0.45. When performing a binary classification task, the classifier describes their prediction of the class of data as `positive' or `negative'. The terms `true' and `false' are then used to describe whether these predictions correspond to the correct classes or `observations'.

\subsection{Machine learning algorithms}
\label{sec:MLalgo}
Our aim with this study was to train ML algorithms to predict whether a given transient is an SN Ic-BL. We used binary classification, as there are only two possible outcomes to the question: yes, the transient is an SN Ic-BL or \textit{no}, the transient is not an SN Ic-BL. A true positive (TP) corresponds to a transient predicted and observed to be an SN Ic-BL; a false positive (FP) refers to a transient predicted to be an SN Ic-BL but observationally confirmed not to be an SN Ic-BL; a true negative (TN) is a transient predicted not to be an SN Ic-BL and confirmed observationally not to be an SN Ic-BL; and finally a false negative (FN) refers to a transient predicted not to be an SN Ic-BL but confirmed observationally to be an SN Ic-BL.

The binary classification task was tested with nine different ML algorithms. Logistic regression is an ML algorithm that falls into the supervised learning category. It is used in binary classifications, where it converts inputs into a probability between zero (false) and one (true) using an S-shaped logistic (sigmoid) function \citep{LR}.The support vector machines algorithm is another example of a supervised ML that searches for the hyperplane that best separates the two possible outcomes (yes or no) \citep{SVM}. The decision trees algorithm consists of a tree-like model of decisions and the consequences of these decisions. The data are divided into smaller groups based on the most significant feature at each tree node \citep{dtree}. The random forest algorithm is an ensemble learning method that constructs multiple decision trees when training a model \citep{randfor}. The AdaBoost algorithm is also an ensemble learning method. It uses decision trees as a weak estimator and initially assigns equal weights to all data points. It then assigns a stronger weighting to the data points that are wrongly classified. This repeats until a low error in the model is obtained \citep{adaboost}. Naive Bayes is a supervised ML algorithm that applies Bayes's theorem, which makes the assumption that all predictors in the model are conditionally independent \citep{naivebayes}. The K-nearest neighbour test is a non-parametric supervised learning algorithm. The algorithm makes predictions by calculating the distance between all the data points in the training sample and those in the testing sample \citep{kkn}. Multi-layer perceptron (MLP) classifier algorithms use underlying neural networks to perform classifications where input data points are mapped to a set of appropriate outputs \citep{mlp}. Quadratic discriminant algorithms are supervised ML algorithms that assume that the observed data \textcolor{red}{are} drawn from a Gaussian distribution with class-specific mean vectors and class-specific covariance matrices. These algorithms use a quadratic score function to assign observations to classes \citep{qd}.

\subsection{Machine learning implementation}
The ML models were trained and tested using nine parameters: three magnitudes, two time differences between consecutive magnitude points, the time difference between the first and third magnitude points, two magnitude rates, and the second derivative of the magnitude rates. The final results were obtained through the median output from 500 iterations of each ML algorithm considered. Transients were randomly shuffled and split into 75\% from training and 25\% for testing. Separate small validation datasets were also created. The Ic-BL data points were evenly distributed between the testing and training sets to prevent class imbalance. This approach ensured that each iteration used different combinations of transients for training and testing.

There were 265 SN Ic-BL data points in the dataset, including multiple filters for some individual SNe Ic-BL. Due to the rarity of SNe Ic-BL, there is a much larger pool of non-Ic-BL transients available to train the ML models.

At each ML run, accuracy, precision, recall, and F1 scores were calculated to evaluate each model’s performance in the training and validation. The following formulae were used for the calculation of these metrics:

\begin{equation}
\mathrm{Accuracy} = \frac{\mathrm{TP} + \mathrm{TN}}{\mathrm{TP} + \mathrm{TN} + \mathrm{FP} + \mathrm{FN}},
\end{equation}

\begin{equation}
\mathrm{Precision} = \frac{\mathrm{TP}}{\mathrm{TP} + \mathrm{FP}},
\end{equation}

\begin{equation}
\mathrm{Recall} = \frac{\mathrm{TP}}{\mathrm{TP} + \mathrm{FN}},
\end{equation}

\begin{equation}
\mathrm{F1} = 2 \times \frac{\mathrm{Precision} \times \mathrm{Recall}}{\mathrm{Precision} + \mathrm{Recall}}.
\end{equation}

The model hyperparameters were then tuned using the inbuilt Scikit-learn function GridSearchCV \citep{scikit-learn}. This function explores combinations of different hyperparameters provided by the user in a predefined dictionary and evaluates the model's performance using the cross-validation method for each hyperparameter combination. By default, the function chooses the hyperparameter combination that optimises the accuracy score. However, in our case, we created a simple custom scorer that prioritised precision and F1 score equally, as we observed that prioritising just precision led to a large increase in the number of SNe Ic-BL being missed. By prioritising both precision and F1 score equally, the function returned the hyperparameters that minimise the FP rate while also reducing the number of `missed' SNe Ic-BL.

\subsection{Addressing class imbalance}
Class imbalance proves to be a significant issue when trying to implement ML for classification purposes. According to TNS, only about 0.8\% of SNe classified per annum on average are SNe Ic-BL. This would create a massive class imbalance if one were to implement this in training a model with ML. A total of 265 SN Ic-BL datasets (minority class) were available to us, in contrast to the thousands of available non-Ic-BL transients. Creating a dataset with SN Ic-BL making up just 0.8\% of the total dataset would lead to the model becoming biased towards the majority class, and consequently the model would be very good at identifying what is not an SN Ic-BL and would be more likely to discard all the "true" SNe Ic-BL. The dataset thus had to be scaled and resampled such that the SNe Ic-BL data points make up a larger portion of the ML dataset. 

We therefore performed an investigation of class imbalance effects through separate ML runs: one in which the number of SNe Ic-BL and non-SNe Ic-BL were balanced in the training and another in which a slight class imbalance was introduced. Table \ref{numbers_in_runs} shows the breakdown of the numbers of each transient for each run of the ML. For the 50-50 balanced case, a total of 488 transients were used in training the ML (244 yes and 244 no). For the imbalanced dataset investigation, the number of non-Ic-BL transients was increased to 568. This allowed the SN Ic-BL population to make up approximately 30\% of the total dataset. All of these cases were also performed with SNe Ia as a control.

\begin{table}[h!]
\renewcommand{\arraystretch}{1.1}
\centering
\caption{
Breakdown of the numbers of transients used for training and evaluation of the machine-learning models.}
\small
\begin{tabular}{l c c c c}
\hline
Run & Dataset & Yes & No & Total \\
\hline
5050 Ia      & Train/Test & 244 & 244 & 488 \\
5050 Ic-BL   & Train/Test & 244 & 244 & 488 \\
5050 Ia      & Real-life  & 50  & 50  & 100 \\
5050 Ic-BL   & Real-life  & 22  & 43  & 65  \\
\hline
7030 Ia      & Train/Test & 244 & 568 & 812 \\
7030 Ic-BL   & Train/Test & 244 & 568 & 812 \\
7030 Ia      & Real-life  & 50  & 50  & 100 \\
7030 Ic-BL   & Real-life  & 22  & 43  & 65  \\
\hline
\end{tabular}
\tablefoot{The 'real-life' dataset corresponds to an entirely unseen sample used for the final validation of the ML models.}
\label{numbers_in_runs}
\end{table}

\section{Results}
\label{sec:results}

Using the nine-parameter space defined in this study, the algorithms described in Sect.~\ref{sec:MLalgo} were employed for each SN. Due to the larger number of SNe Ia available, the classification method was also applied to SN Ia to test the performance of this new approach in comparison to the more limited sample of SNe Ic-BL. In each case, the ML was run 500 times, randomly shuffling the transients to obtain a median result. All the tables summarising the different metrics measured in each case are shown in Appendix~\ref{app:tables}.

\subsection{50-50 distribution}
Figure~\ref{im_5050} shows the median performance of each model across the 500 runs when aiming to identify SN Ia or SNe Ic-BL. In both cases, the model that performed the best at minimising FP and maximising TP counts was the random forest algorithm. When looking at tables \ref{iatrain5050} and \ref{icbltrain5050}, we noted that the precision score for both ML cases was quite good and consistent with each other (0.76 in the case of SNe Ia versus 0.71 for SNe Ic-BL).

On the other hand, the recall scores seem to differ between the two cases, with the SN Ia run returning a much higher value (0.77 for SNe Ia versus 0.44 for SNe Ic-BL), which suggests that the ML approach used in this study seems to fail to identify a larger portion of relevant SN Ic-BL cases. This is likely caused by the current lack of a large SN Ic-BL population.

From the training and validation data, we observed that the models perform well in both cases. To put these models to the test, we created a real-life test scenario with a completely new sample of SNe. This allowed us to simulate how the models would be rolled out in real-life and explore how well they generalise to new data.

As shown in figure~\ref{real_5050}, random forest was again observed to perform the best in each case for the real-life testing scenario. While the SN Ia case seemed to keep performing well, with a precision score of 0.76, we note that the Ic-BL model struggled to generalise to the unseen new data, showing a drop of 46\% in its precision score and a drop of 27\% in its recall score. Tables~\ref{iaval5050} and ~\ref{icblval5050} show the results of all the metrics calculated for each ML algorithm considered for this real-life test scenario.

\begin{figure*}

    \centering
        \begin{subfigure}[t]{0.35\textwidth}
            \centering
            \includegraphics[width=\textwidth]{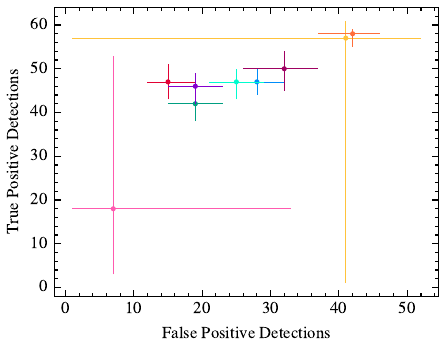}
            \caption{SN Ia}
  
        \end{subfigure}%
        \hspace{0.05\textwidth}
        \begin{subfigure}[t]{0.44\textwidth}
            \centering
            \includegraphics[width=\textwidth]{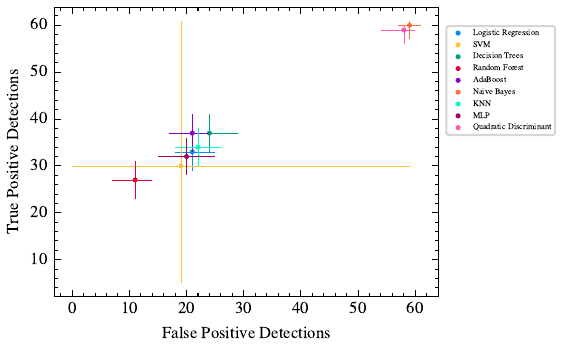}
            \caption{SN Ic-BL}
        \end{subfigure}
        \caption{Median results from the training and validation sample after 500 runs using the 50-50 distribution of `yes' and `no'.}
    \label{im_5050}
\end{figure*}

\begin{figure*}

    \centering
        \begin{subfigure}[t]{0.35\textwidth}
            \centering
            \includegraphics[width=\textwidth]{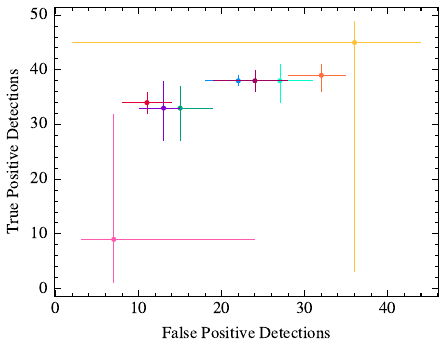}
            \caption{SN Ia}

        \end{subfigure}%
        \hspace{0.05\textwidth}
        \begin{subfigure}[t]{0.44\textwidth}
            \centering
            \includegraphics[width=\textwidth]{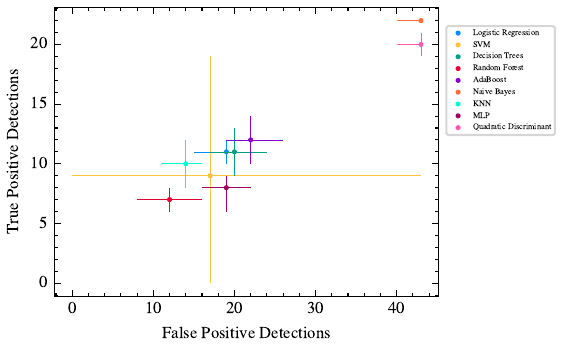}
            \caption{SN Ic-BL}

        \end{subfigure}
        \caption{Median results from the real-life testing scenario from the 500 runs using the 50-50 distribution model.}
    \label{real_5050}
\end{figure*}

\subsection{70-30 distribution}
The potential impact of our approach using an imbalanced sample was tested with a 70-30 distribution following the same methodology as described in the balanced scenario. Here, the SN Ia results dropped slightly across all scores when the distribution of the dataset was imbalanced, as can be seen in Table~\ref{ia7030_train}. In contrast, the SN Ic-Bl model appeared to benefit from a slight imbalance in the dataset. The precision value in Table~\ref{icbl7030_train} for the random forest model increased to 0.83 in the training and validation dataset, thus increasing the reliability of the model. This, however, came at a cost, as the recall scores for the random forest model dropped to 0.16. 

In both cases, the models for SNe Ia and SNe Ic-BL were observed to generalise well in the real-life dataset scenario, as shown in figure~\ref{7030val}. The SN Ia random forest model still performed relatively well, with a precision score of 0.80, as shown in table~\ref{ia_7030_val}. The results for the SN Ic-BL dataset found in table~\ref{icbl7030_val} return a precision score of one for the random forest model, meaning, on average, no FPs were returned when the model was tested on this particular dataset. The logistic regression algorithm also returned quite good scores in the SN Ic-BL model run, with a precision score of one. 

\begin{figure*}
    
    \centering
    \begin{subfigure}[t]{0.35\textwidth}
        \centering
        
        \includegraphics[width=\textwidth]{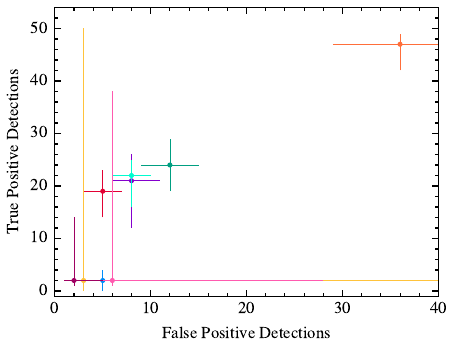}
        \caption{SN Ia}
         \label{7030valia}
    \end{subfigure}%
        \hspace{0.05\textwidth}
    \begin{subfigure}[t]{0.44\textwidth}
        \centering
        \includegraphics[width=\textwidth]{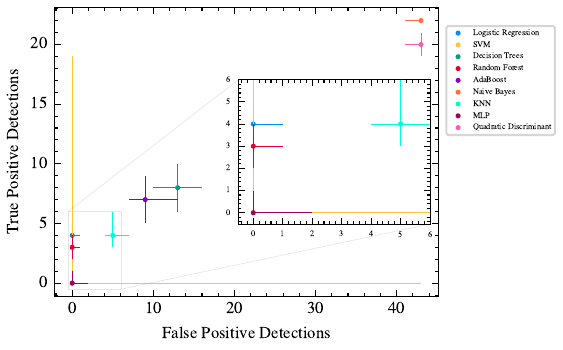}
        \caption{SN Ic-BL}
        \label{7030valicbl}
    \end{subfigure}
    \caption{Median results from the real-life dataset testing scenario from the 500 runs using the 70-30 distribution model.}
    \label{7030val}
\end{figure*}

\section{Conclusion}
\label{sec:conclusions}

This study demonstrates the ability of ML algorithms to identify trends in SN data despite limited photometric information. When training the ML on SNe Ic-BL, we observed that the model struggles to effectively generalise to unseen data, especially in the balanced 50-50 training case. Adjusting the training dataset to the 70-30 split between SNe Ic-BL and non-SNe Ic-BL showed an improvement in generalisation. Exposing the model to more non-SN Ic-BL examples seems to help refine the classification criteria, thus reducing the rate of FPs. This, however, comes at a cost, as the model misses a larger number of true SNe Ic-BL, resulting in a lower recall score.

In contrast, the SN Ia dataset performs best in the 50-50 training dataset distribution, though the scores dropped only slightly in the 70-30 model run. That being said, the model still performs decently well and generalises satisfactorily to unseen data in both cases. The greater abundance and the availability of better quality photometric data of SNe Ia allow the model to recognise better patterns and underlying distributions within the data, leading the model to perform well in the training and generalise well to the real-life dataset scenario.

\subsection{Machine learning potential with improved datasets}
It is important to note that the SN Ic-BL dataset is limited to only 136 unique transients with good quality data, which proves to be a key challenge for accurate classifications. The quality of the few SN Ic-BL data points is also lacking in comparison to that of SNe Ia. This disparity in good-quality data may explain the difference in the performance of the models in both SN classes. With the collection of higher-quality SN Ic-BL photometric data over time, the performance of the SN Ic-BL models should improve and more accurately classify transients. Figure~\ref{f1all} shows a direct comparison of the F1 score between the two SN samples, providing a good projection of the potential of the ML approach we present as the number of good-quality SNe Ic-BL increases.

\begin{figure}[h!]
    \centering
    \includegraphics[width=0.34\textwidth]{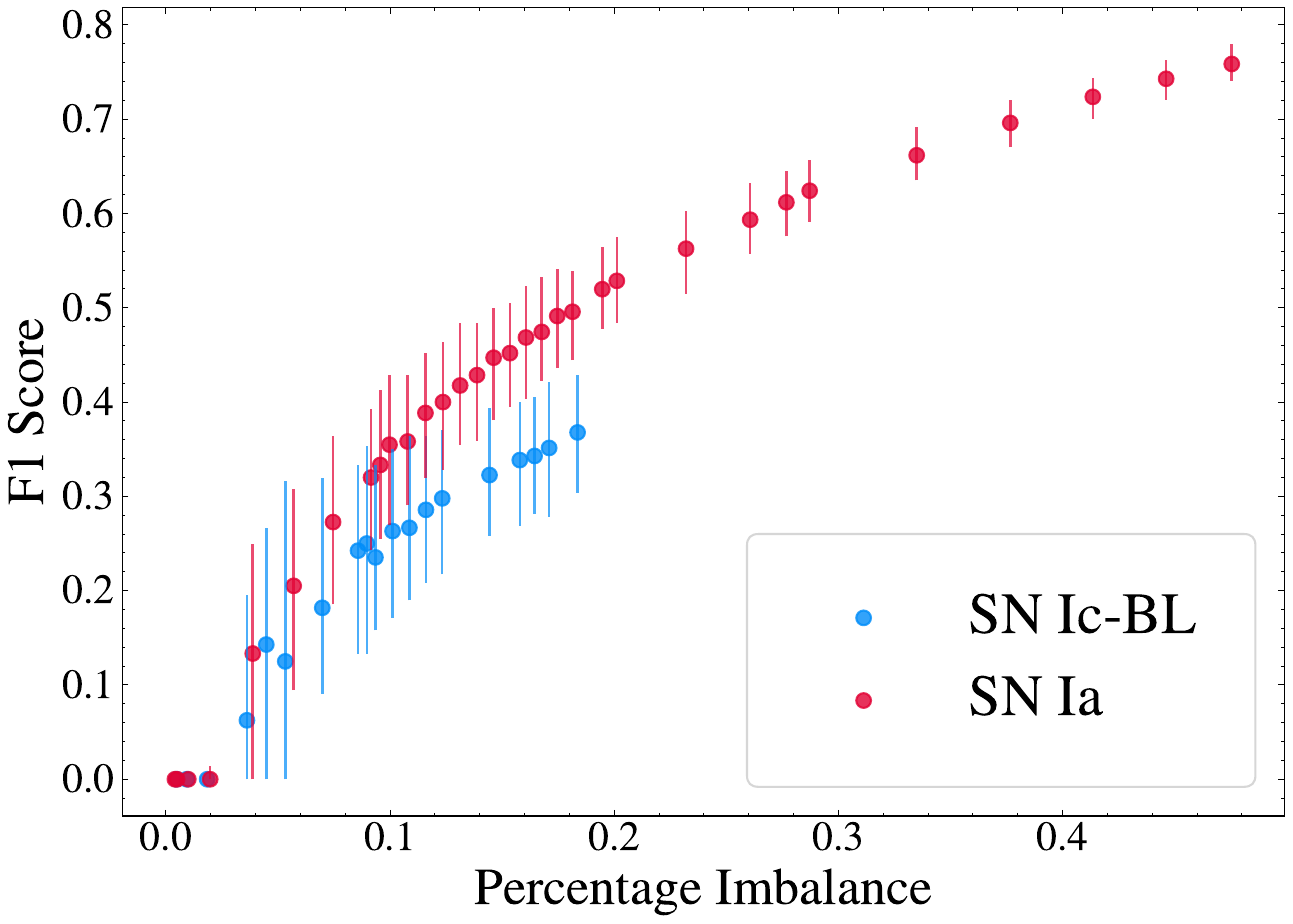}
    \caption{F1 scores from random forest models versus imbalance in the dataset. The percentage imbalance refers to the ratio of the number of the target (yes) SNe to the number of other (no) SNe in the dataset (i.e. the class imbalance in a given dataset).}
    \label{f1all}
\end{figure}

Figure~\ref{f1all} clearly illustrates the random forest model's performance, as the class imbalance present in the dataset changes. The SN Ia dataset contains 1067 `no' transients. The class imbalance reduces as the number of `yes' transients increases towards 1000, resulting in the F1 score gradually improving. At this point, the model misses fewer of the `yes' transients and makes fewer mistakes in classifying them. As more SNe Ic-BL are discovered and data are incorporated into the model, the performance of these ML models is expected to improve significantly.

For this study, only a very small unseen dataset could be spared to perform the real-life scenario test in the case of the SN Ic-BL sample. As significantly more data were available to test the behaviour of this method on SNe Ia, additional data from a new SN in 2024 were also considered. The results were significantly worse for the SN Ia model when this extra validation dataset was used, as seen in Table~\ref{24unseen}. This discrepancy in scores likely arises from misclassifications due to the inclusion of recently classified transients that may have been misclassified. This suggests that the model's ability to generalise to unseen data is highly dependent on the specific characteristics of the unseen dataset.

\begin{table}
\renewcommand{\arraystretch}{1.25}
  \centering
    \caption{Median results from the ML identification of SNe Ia from the 2024 unseen dataset.}
    \small
    \begin{tabular}{l c c c }  
    \hline
       Accuracy  & Precision  & Recall & F1\_Score \\
       \hline
        0.59 & 0.57 & 0.67 & 0.62 \\
        \hline
    \end{tabular}
 
    \label{24unseen}
\end{table}

Current models are still missing and failing to observe the early stages of a significant number of SNe Ic-BL.  In 2022, 14 SNe Ic-BL were detected with good-quality light curves. From our initial order of magnitude approximation, if we assume that $\sim$150 SN Ic-BL events occur annually, the current capability of detecting SNe Ic-BL based on the number of SNe Ic-BL detected in 2022 is 9.3\%. The results from the real-life dataset test of the model trained on the 70-30 distribution presented in figure \ref{7030val} and table \ref{icbl7030_val} identify a median TP value of three SNe Ic-BL out of 22 and no FPs. This indicates that our proposed approach has the potential to identify upwards of  13.6\% of the SN Ic-BL population, which is an improvement from the 2022 transients. This would mean that the ML would be able to detect one in ten SNe Ic-BL per year. The availability of this new method of early SNe Ic-BL identification can open the possibility of dedicated campaigns specifically designed to classify Ic-BLs and improve their light curve monitoring, which would result in a much higher quality dataset.

\subsection{Future investigations and implementations}
With the Vera C. Rubin Observatory now operational and the Legacy Survey of Space and Time (LSST) scheduled to commence in 2026, we anticipate a substantial increase in the discovery and spectroscopic classification of SNe Ic-BL. The integration of these ML models with real-time observations, such as those from LSST, will also substantially improve the quality of SNe data. 

The LSST will issue alerts for new transients, and brokers such as \texttt{ALeRCE} will ingest these alerts and provide the stamp classifications for each object.\footnote{\url{https://rubinobservatory.org/for-scientists/data-products/alerts-and-brokers}} This will provide a starting point for potential SN Ic-BL candidates. The LSST's Wide Fast Deep (WFD) survey will revisit the same point in the sky at a cadence of 2-4 days, enabling the acquisition of at least two early photometric data points within the first week following explosion.\footnote{\url{https://survey-strategy.lsst.io/baseline/wfd.html}} Our ML model works well with just the first three data points. Thus, coordinated observations with complementary facilities could secure a third data point and would allow for the ML model to be implemented at very early times, providing robust justification for spectroscopic follow-up of suitable candidates. Furthermore, LSST’s deep imaging capabilities will provide a 5$\sigma$ depth of 24.44 AB mag in the \textit{g}-band and 23.98 AB mag in the \textit{r}-band in a single exposure, enabling the coverage of earlier and fainter photometric points \citep{Bianco_22_LSST_im_depth}.

Thus, these new surveys will play a pivotal role in capturing the earliest phases of SN Ic-BL evolution. Early-time detections will provide sufficient motivation for spectroscopic follow-up closer to the SN trigger time, allowing for valuable investigations of early-time spectroscopic evolution. Moreover, early-time observations will facilitate the production of high-quality light curves with well-sampled early-rising data, which is currently lacking in SN Ic-BL photometric datasets. Advances in the quality and quantity of these datasets will be key to gaining an understanding of the properties of the progenitor and bridging the critical gaps in our knowledge of the GRB-SN phenomena.

\begin{acknowledgements}
The authors would like to thank the anonymous referee for their comments that helped improve this paper. LC and AMC acknowledge the support of the Irish Research Council Postgraduate Scholarship No GOIPG/2022/1008. GF and AMC acknowledge the support of the UCD Ad Astra program. JF and AMC acknowledge support from the European Space Agency (PRODEX) (Grant No. 4000138314).

LC would like to thank the Berkeley SETI Research Center at UC Berkeley and everyone at Hat Creek Observatory, California, for their kindness and hospitality in providing a workspace during a three-week visit in September 2024, where much of this paper was written. 

LC would like to thank Owen A. Johnson for their insightful comments and discussions.

\end{acknowledgements}

\bibliographystyle{aa}
\bibliography{references}

\clearpage
\appendix

\twocolumn[
 
    \section{Additional tables}
    \label{app:tables}
    Tables showing the different metrics measured in each of the training and testing datasets for each SN case and for each of the ML
    algorithms models considered. Each value for Accuracy, Precision, Recall and F1 score are the median value of a total of 500 runs. The
    associated uncertainties are estimated using the 16th and 84th percentiles of the distribution, measured relative to the median. In
    some cases where the median value was close to the boundary of the allowed range (0,1), the percentile-based uncertainties can
    extend beyond this range. In such instances, the uncertainties are truncated such that the reported values remain within the physically
    allowed bounds.

    \vspace{4em}]

\begin{table}
    
    \centering
    \caption{Results of the validation done from a training set with a 50-50 distribution to identify SN Ia.}
    \label{iatrain5050}
    \tiny
        \begin{tabular}{l c c c c}  
        \hline
        Model & Accuracy & Precision & Recall & F1 Score\\ \hline
        Logistic Regression & 
    0.66 \raisebox{0.5ex}{\tiny$^{+0.04}_{-0.04}$} & 
    0.63 \raisebox{0.5ex}{\tiny$^{+0.04}_{-0.04}$} & 
    0.77 \raisebox{0.5ex}{\tiny$^{+0.05}_{-0.05}$} & 
    0.69 \raisebox{0.5ex}{\tiny$^{+0.03}_{-0.04}$} \\ 
    SVM & 
    0.63 \raisebox{0.5ex}{\tiny$^{+0.06}_{-0.06}$} & 
    0.58 \raisebox{0.5ex}{\tiny$^{+0.07}_{-0.21}$} & 
    0.93 \raisebox{0.5ex}{\tiny$^{+0.07}_{-0.90}$} & 
    0.71 \raisebox{0.5ex}{\tiny$^{+0.03}_{-0.65}$} \\ 
    Decision Trees & 
    0.68 \raisebox{0.5ex}{\tiny$^{+0.04}_{-0.04}$} & 
    0.68 \raisebox{0.5ex}{\tiny$^{+0.05}_{-0.04}$} & 
    0.69 \raisebox{0.5ex}{\tiny$^{+0.07}_{-0.07}$} & 
    0.68 \raisebox{0.5ex}{\tiny$^{+0.04}_{-0.05}$} \\ 
    Random Forest & 
    0.76 \raisebox{0.5ex}{\tiny$^{+0.03}_{-0.04}$} & 
    0.76 \raisebox{0.5ex}{\tiny$^{+0.04}_{-0.04}$} & 
    0.77 \raisebox{0.5ex}{\tiny$^{+0.07}_{-0.07}$} & 
    0.76 \raisebox{0.5ex}{\tiny$^{+0.03}_{-0.04}$} \\ 
    AdaBoost & 
    0.72 \raisebox{0.5ex}{\tiny$^{+0.04}_{-0.03}$} & 
    0.71 \raisebox{0.5ex}{\tiny$^{+0.04}_{-0.04}$} & 
    0.76 \raisebox{0.5ex}{\tiny$^{+0.07}_{-0.06}$} & 
    0.74 \raisebox{0.5ex}{\tiny$^{+0.04}_{-0.04}$} \\ 
    Naive Bayes & 
    0.62 \raisebox{0.5ex}{\tiny$^{+0.04}_{-0.03}$} & 
    0.58 \raisebox{0.5ex}{\tiny$^{+0.03}_{-0.02}$} & 
    0.95 \raisebox{0.5ex}{\tiny$^{+0.02}_{-0.05}$} & 
    0.71 \raisebox{0.5ex}{\tiny$^{+0.02}_{-0.02}$} \\ 
    KNN & 
    0.67 \raisebox{0.5ex}{\tiny$^{+0.04}_{-0.04}$} & 
    0.65 \raisebox{0.5ex}{\tiny$^{+0.04}_{-0.04}$} & 
    0.77 \raisebox{0.5ex}{\tiny$^{+0.05}_{-0.07}$} & 
    0.70 \raisebox{0.5ex}{\tiny$^{+0.04}_{-0.04}$} \\ 
    MLP & 
    0.65 \raisebox{0.5ex}{\tiny$^{+0.04}_{-0.04}$} & 
    0.61 \raisebox{0.5ex}{\tiny$^{+0.03}_{-0.03}$} & 
    0.82 \raisebox{0.5ex}{\tiny$^{+0.07}_{-0.08}$} & 
    0.70 \raisebox{0.5ex}{\tiny$^{+0.04}_{-0.04}$} \\ 
    Quadratic Discriminant & 
    0.58 \raisebox{0.5ex}{\tiny$^{+0.11}_{-0.07}$} & 
    0.71 \raisebox{0.5ex}{\tiny$^{+0.18}_{-0.08}$} & 
    0.30 \raisebox{0.5ex}{\tiny$^{+0.60}_{-0.25}$} & 
    0.42 \raisebox{0.5ex}{\tiny$^{+0.31}_{-0.33}$} \\ 
\hline

        \end{tabular}
 
\end{table}

\begin{table}
    
    \centering
    \caption{Results of the validation done from a training set with a 50-50 distribution to identify SNe Ic-BL.}
    \label{icbltrain5050}
    \tiny
        \begin{tabular}{l c c c c}  
        \hline
        Model & Accuracy & Precision & Recall & F1 Score\\ \hline
        Logistic Regression & 
    0.60 \raisebox{0.5ex}{\tiny$^{+0.04}_{-0.04}$} & 
    0.60\raisebox{0.5ex}{\tiny$^{+0.04}_{-0.05}$} & 
    0.54 \raisebox{0.5ex}{\tiny$^{+0.07}_{-0.07}$} & 
    0.57 \raisebox{0.5ex}{\tiny$^{+0.05}_{-0.05}$} \\ 
    SVM & 
    0.59 \raisebox{0.5ex}{\tiny$^{+0.02}_{-0.03}$} & 
    0.61 \raisebox{0.5ex}{\tiny$^{+0.04}_{-0.39}$} & 
    0.49 \raisebox{0.5ex}{\tiny$^{+0.51}_{-0.21}$} & 
    0.55 \raisebox{0.5ex}{\tiny$^{+0.49}_{-0.03}$} \\ 
    Decision Trees & 
    0.60 \raisebox{0.5ex}{\tiny$^{+0.04}_{-0.05}$} & 
    0.60 \raisebox{0.5ex}{\tiny$^{+0.05}_{-0.05}$} & 
    0.61 \raisebox{0.5ex}{\tiny$^{+0.07}_{-0.07}$} & 
    0.60 \raisebox{0.5ex}{\tiny$^{+0.05}_{-0.05}$} \\ 
    Random Forest & 
    0.63 \raisebox{0.5ex}{\tiny$^{+0.04}_{-0.03}$} & 
    0.71 \raisebox{0.5ex}{\tiny$^{+0.07}_{-0.06}$} & 
    0.44 \raisebox{0.5ex}{\tiny$^{+0.08}_{-0.07}$} & 
    0.55 \raisebox{0.5ex}{\tiny$^{+0.06}_{-0.06}$} \\ 
    AdaBoost & 
    0.63 \raisebox{0.5ex}{\tiny$^{+0.04}_{-0.04}$} & 
    0.64 \raisebox{0.5ex}{\tiny$^{+0.05}_{-0.05}$} & 
    0.61 \raisebox{0.5ex}{\tiny$^{+0.07}_{-0.05}$} & 
    0.62 \raisebox{0.5ex}{\tiny$^{+0.05}_{-0.04}$} \\ 
    Naive Bayes & 
    0.50 \raisebox{0.5ex}{\tiny$^{+0.02}_{-0.02}$} & 
    0.50 \raisebox{0.5ex}{\tiny$^{+0.01}_{-0.01}$} & 
    0.98 \raisebox{0.5ex}{\tiny$^{+0.02}_{-0.02}$} & 
    0.66 \raisebox{0.5ex}{\tiny$^{+0.02}_{-0.01}$} \\ 
    KNN & 
    0.60 \raisebox{0.5ex}{\tiny$^{+0.04}_{-0.04}$} & 
    0.61 \raisebox{0.5ex}{\tiny$^{+0.05}_{-0.05}$} & 
    0.56 \raisebox{0.5ex}{\tiny$^{+0.07}_{-0.07}$} & 
    0.58 \raisebox{0.5ex}{\tiny$^{+0.05}_{-0.05}$} \\ 
    MLP & 
    0.60 \raisebox{0.5ex}{\tiny$^{+0.04}_{-0.05}$} & 
    0.62 \raisebox{0.5ex}{\tiny$^{+0.05}_{-0.05}$} & 
    0.52 \raisebox{0.5ex}{\tiny$^{+0.08}_{-0.07}$} & 
    0.57 \raisebox{0.5ex}{\tiny$^{+0.06}_{-0.05}$} \\ 
    Quadratic Discriminant & 
    0.51 \raisebox{0.5ex}{\tiny$^{+0.02}_{-0.03}$} & 
    0.50 \raisebox{0.5ex}{\tiny$^{+0.01}_{-0.02}$} & 
    0.95 \raisebox{0.5ex}{\tiny$^{+0.05}_{-0.03}$} & 
    0.66 \raisebox{0.5ex}{\tiny$^{+0.02}_{-0.02}$} \\ 
\hline

        \end{tabular}
 
\end{table}

\begin{table}
    \centering
    \caption{Results of the real-life dataset test on the 50-50 distribution trained models to identify SN Ia.}
    \label{iaval5050}
    \tiny
        \begin{tabular}{l c c c c}  
        \hline
        Model & Accuracy & Precision & Recall & F1 Score\\ \hline
        Logistic Regression & 
    0.66 \raisebox{0.5ex}{\tiny$^{+0.05}_{-0.04}$} & 
0.63 \raisebox{0.5ex}{\tiny$^{+0.05}_{-0.04}$} & 
0.76 \raisebox{0.5ex}{\tiny$^{+0.02}_{-0.02}$} & 
0.69 \raisebox{0.5ex}{\tiny$^{+0.03}_{-0.03}$} \\ 
SVM & 
0.55 \raisebox{0.5ex}{\tiny$^{+0.06}_{-0.04}$} & 
0.58 \raisebox{0.5ex}{\tiny$^{+0.20}_{-0.04}$} & 
0.90 \raisebox{0.5ex}{\tiny$^{+0.08}_{-0.84}$} & 
0.66 \raisebox{0.5ex}{\tiny$^{+0.04}_{-0.55}$} \\ 
Decision Trees & 
0.68 \raisebox{0.5ex}{\tiny$^{+0.05}_{-0.07}$} & 
0.68 \raisebox{0.5ex}{\tiny$^{+0.06}_{-0.06}$} & 
0.66 \raisebox{0.5ex}{\tiny$^{+0.08}_{-0.12}$} & 
0.68 \raisebox{0.5ex}{\tiny$^{+0.05}_{-0.09}$} \\ 
Random Forest & 
0.74 \raisebox{0.5ex}{\tiny$^{+0.03}_{-0.04}$} & 
0.76 \raisebox{0.5ex}{\tiny$^{+0.05}_{-0.05}$} & 
0.69 \raisebox{0.5ex}{\tiny$^{+0.05}_{-0.05}$} & 
0.72 \raisebox{0.5ex}{\tiny$^{+0.04}_{-0.04}$} \\ 
AdaBoost & 
0.71 \raisebox{0.5ex}{\tiny$^{+0.05}_{-0.08}$} & 
0.72 \raisebox{0.5ex}{\tiny$^{+0.05}_{-0.06}$} & 
0.66 \raisebox{0.5ex}{\tiny$^{+0.10}_{-0.12}$} & 
0.70 \raisebox{0.5ex}{\tiny$^{+0.06}_{-0.10}$} \\ 
Naive Bayes & 
0.57 \raisebox{0.5ex}{\tiny$^{+0.03}_{-0.02}$} & 
0.55 \raisebox{0.5ex}{\tiny$^{+0.02}_{-0.02}$} & 
0.78 \raisebox{0.5ex}{\tiny$^{+0.04}_{-0.06}$} & 
0.64 \raisebox{0.5ex}{\tiny$^{+0.02}_{-0.02}$} \\ 
KNN & 
0.60 \raisebox{0.5ex}{\tiny$^{+0.05}_{-0.04}$} & 
0.58 \raisebox{0.5ex}{\tiny$^{+0.03}_{-0.04}$} & 
0.76 \raisebox{0.5ex}{\tiny$^{+0.06}_{-0.08}$} & 
0.65 \raisebox{0.5ex}{\tiny$^{+0.04}_{-0.04}$} \\ 
MLP & 
0.64 \raisebox{0.5ex}{\tiny$^{+0.04}_{-0.03}$} & 
0.61 \raisebox{0.5ex}{\tiny$^{+0.04}_{-0.03}$} & 
0.76 \raisebox{0.5ex}{\tiny$^{+0.04}_{-0.04}$} & 
0.68 \raisebox{0.5ex}{\tiny$^{+0.03}_{-0.03}$} \\ 
Quadratic Discriminant & 
0.51 \raisebox{0.5ex}{\tiny$^{+0.06}_{-0.04}$} & 
0.56 \raisebox{0.5ex}{\tiny$^{+0.05}_{-0.32}$} & 
0.18 \raisebox{0.5ex}{\tiny$^{+0.46}_{-0.16}$} & 
0.27 \raisebox{0.5ex}{\tiny$^{+0.33}_{-0.24}$} \\
\hline

        \end{tabular}
 
\end{table}

\begin{table}
    \centering
    \caption{Results of the real-life dataset test on the 50-50 distribution trained models to identify SNe Ic-BL.}
    \label{icblval5050}
    \tiny
    \begin{tabular}{l c c c c}  
    \hline
        Model & Accuracy  & Precision  & Recall & F1\_Score \\
        \hline
        Logistic Regression & 
        0.52 \raisebox{0.5ex}{\tiny$^{+0.08}_{-0.08}$} & 
0.35 \raisebox{0.5ex}{\tiny$^{+0.05}_{-0.06}$} & 
0.50 \raisebox{0.5ex}{\tiny$^{+0.05}_{-0.05}$} & 
0.41 \raisebox{0.5ex}{\tiny$^{+0.04}_{-0.04}$} \\
SVM & 
0.55 \raisebox{0.5ex}{\tiny$^{+0.24}_{-0.12}$} & 
0.34 \raisebox{0.5ex}{\tiny$^{+0.32}_{-0.12}$} & 
0.43 \raisebox{0.5ex}{\tiny$^{+0.43}_{-0.48}$} & 
0.37 \raisebox{0.5ex}{\tiny$^{+0.27}_{-0.10}$} \\
Decision Trees & 
0.52 \raisebox{0.5ex}{\tiny$^{+0.08}_{-0.06}$} & 
0.35 \raisebox{0.5ex}{\tiny$^{+0.06}_{-0.06}$} & 
0.50 \raisebox{0.5ex}{\tiny$^{+0.09}_{-0.09}$} & 
0.42 \raisebox{0.5ex}{\tiny$^{+0.08}_{-0.07}$} \\
Random Forest & 
0.60 \raisebox{0.5ex}{\tiny$^{+0.05}_{-0.06}$} & 
0.38 \raisebox{0.5ex}{\tiny$^{+0.06}_{-0.11}$} & 
0.32 \raisebox{0.5ex}{\tiny$^{+0.05}_{-0.05}$} & 
0.36 \raisebox{0.5ex}{\tiny$^{+0.04}_{-0.05}$} \\
AdaBoost & 
0.49 \raisebox{0.5ex}{\tiny$^{+0.06}_{-0.06}$} & 
0.34 \raisebox{0.5ex}{\tiny$^{+0.06}_{-0.06}$} & 
0.55 \raisebox{0.5ex}{\tiny$^{+0.09}_{-0.09}$} & 
0.42 \raisebox{0.5ex}{\tiny$^{+0.07}_{-0.06}$} \\
Naive Bayes & 
0.34 \raisebox{0.5ex}{\tiny$^{+0.00}_{-0.05}$} & 
0.34 \raisebox{0.5ex}{\tiny$^{+0.00}_{-0.02}$} & 
1.00 \raisebox{0.5ex}{\tiny$^{+0.00}_{-0.00}$} & 
0.51 \raisebox{0.5ex}{\tiny$^{+0.00}_{-0.01}$} \\
KNN & 
0.60 \raisebox{0.5ex}{\tiny$^{+0.05}_{-0.05}$} & 
0.42 \raisebox{0.5ex}{\tiny$^{+0.06}_{-0.06}$} & 
0.45 \raisebox{0.5ex}{\tiny$^{+0.09}_{-0.09}$} & 
0.43 \raisebox{0.5ex}{\tiny$^{+0.06}_{-0.06}$} \\
MLP & 
0.48 \raisebox{0.5ex}{\tiny$^{+0.03}_{-0.05}$} & 
0.29 \raisebox{0.5ex}{\tiny$^{+0.05}_{-0.03}$} & 
0.32 \raisebox{0.5ex}{\tiny$^{+0.05}_{-0.09}$} & 
0.31 \raisebox{0.5ex}{\tiny$^{+0.06}_{-0.04}$} \\
Quadratic Discriminant & 
0.32 \raisebox{0.5ex}{\tiny$^{+0.02}_{-0.03}$} & 
0.32 \raisebox{0.5ex}{\tiny$^{+0.01}_{-0.02}$} & 
0.91 \raisebox{0.5ex}{\tiny$^{+0.05}_{-0.05}$} & 
0.47 \raisebox{0.5ex}{\tiny$^{+0.02}_{-0.03}$} \\
        \hline
    \end{tabular}
\end{table}

\begin{table}
    \centering
    \caption{Results of the validation done from a training set with a 70-30 distribution to identify SNe Ia.}
    \label{ia7030_train}
    \tiny
    \begin{tabular}{l c c c c}
    \hline
        Model & Accuracy & Precision & Recall & F1\_Score \\
    \hline
     Logistic Regression &
        0.68 \raisebox{0.5ex}{\tiny$^{+0.01}_{-0.02}$} &
        0.29 \raisebox{0.5ex}{\tiny$^{+0.14}_{-0.12}$} &
        0.05 \raisebox{0.5ex}{\tiny$^{+0.03}_{-0.03}$} &
        0.08 \raisebox{0.5ex}{\tiny$^{+0.05}_{-0.05}$} \\

        SVM &
        0.68 \raisebox{0.5ex}{\tiny$^{+0.01}_{-0.30}$} &
        0.31 \raisebox{0.5ex}{\tiny$^{+0.09}_{-0.31}$} &
        0.03 \raisebox{0.5ex}{\tiny$^{+0.97}_{-0.03}$} &
        0.06 \raisebox{0.5ex}{\tiny$^{+0.44}_{-0.06}$} \\

        Decision Trees &
        0.71 \raisebox{0.5ex}{\tiny$^{+0.03}_{-0.03}$} &
        0.52 \raisebox{0.5ex}{\tiny$^{+0.05}_{-0.06}$} &
        0.52 \raisebox{0.5ex}{\tiny$^{+0.08}_{-0.07}$} &
        0.53 \raisebox{0.5ex}{\tiny$^{+0.05}_{-0.06}$} \\

        Random Forest &
        0.78 \raisebox{0.5ex}{\tiny$^{+0.02}_{-0.02}$} &
        0.66 \raisebox{0.5ex}{\tiny$^{+0.06}_{-0.05}$} &
        0.52 \raisebox{0.5ex}{\tiny$^{+0.07}_{-0.07}$} &
        0.58 \raisebox{0.5ex}{\tiny$^{+0.05}_{-0.05}$} \\

        AdaBoost &
        0.75 \raisebox{0.5ex}{\tiny$^{+0.02}_{-0.03}$} &
        0.59 \raisebox{0.5ex}{\tiny$^{+0.05}_{-0.06}$} &
        0.56 \raisebox{0.5ex}{\tiny$^{+0.07}_{-0.07}$} &
        0.57 \raisebox{0.5ex}{\tiny$^{+0.05}_{-0.05}$} \\

        Naive Bayes &
        0.49 \raisebox{0.5ex}{\tiny$^{+0.05}_{-0.04}$} &
        0.36 \raisebox{0.5ex}{\tiny$^{+0.02}_{-0.02}$} &
        0.90 \raisebox{0.5ex}{\tiny$^{+0.05}_{-0.07}$} &
        0.51 \raisebox{0.5ex}{\tiny$^{+0.02}_{-0.02}$} \\

        KNN &
        0.69 \raisebox{0.5ex}{\tiny$^{+0.02}_{-0.03}$} &
        0.48 \raisebox{0.5ex}{\tiny$^{+0.05}_{-0.06}$} &
        0.39 \raisebox{0.5ex}{\tiny$^{+0.07}_{-0.07}$} &
        0.43 \raisebox{0.5ex}{\tiny$^{+0.05}_{-0.06}$} \\

        MLP &
        0.70 \raisebox{0.5ex}{\tiny$^{+0.00}_{-0.01}$} &
        0.50 \raisebox{0.5ex}{\tiny$^{+0.26}_{-0.24}$} &
        0.02 \raisebox{0.5ex}{\tiny$^{+0.02}_{-0.02}$} &
        0.03 \raisebox{0.5ex}{\tiny$^{+0.03}_{-0.03}$} \\

        Quadratic Discriminant &
        0.70 \raisebox{0.5ex}{\tiny$^{+0.02}_{-0.09}$} &
        0.50 \raisebox{0.5ex}{\tiny$^{+0.25}_{-0.14}$} &
        0.08 \raisebox{0.5ex}{\tiny$^{+0.72}_{-0.05}$} &
        0.15 \raisebox{0.5ex}{\tiny$^{+0.40}_{-0.09}$} \\
    \hline
    \end{tabular}
\end{table}

\begin{table}
    \centering
    \caption{Results of the validation done from a training set with a 70-30 distribution to identify SNe Ic-BL.}
    \label{icbl7030_train}
    \tiny
    \begin{tabular}{l c c c c}
    \hline
        Model & Accuracy & Precision & Recall & F1\_Score \\
    \hline
        Logistic Regression &
        0.73 \raisebox{0.5ex}{\tiny$^{+0.02}_{-0.01}$} &
        0.72 \raisebox{0.5ex}{\tiny$^{+0.13}_{-0.12}$} &
        0.18 \raisebox{0.5ex}{\tiny$^{+0.05}_{-0.05}$} &
        0.29 \raisebox{0.5ex}{\tiny$^{+0.08}_{-0.07}$} \\

        SVM &
       0.71 \raisebox{0.5ex}{\tiny$^{+0.29}_{-0.01}$} &
        0.86 \raisebox{0.5ex}{\tiny$^{+0.14}_{-0.25}$} &
        0.10 \raisebox{0.5ex}{\tiny$^{+0.05}_{-0.05}$} &
        0.17 \raisebox{0.5ex}{\tiny$^{+0.29}_{-0.08}$} \\

        Decision Trees &
        0.67 \raisebox{0.5ex}{\tiny$^{+0.03}_{-0.03}$} &
        0.44 \raisebox{0.5ex}{\tiny$^{+0.05}_{-0.05}$} &
        0.46 \raisebox{0.5ex}{\tiny$^{+0.07}_{-0.07}$} &
        0.45 \raisebox{0.5ex}{\tiny$^{+0.05}_{-0.05}$} \\

        Random Forest &
        0.74 \raisebox{0.5ex}{\tiny$^{+0.01}_{-0.01}$} &
        0.83 \raisebox{0.5ex}{\tiny$^{+0.13}_{-0.10}$} &
        0.16 \raisebox{0.5ex}{\tiny$^{+0.03}_{-0.05}$} &
        0.28 \raisebox{0.5ex}{\tiny$^{+0.06}_{-0.06}$} \\

        AdaBoost &
        0.73 \raisebox{0.5ex}{\tiny$^{+0.02}_{-0.03}$} &
        0.59 \raisebox{0.5ex}{\tiny$^{+0.07}_{-0.07}$} &
        0.39 \raisebox{0.5ex}{\tiny$^{+0.07}_{-0.05}$} &
        0.47 \raisebox{0.5ex}{\tiny$^{+0.05}_{-0.06}$} \\

        Naive Bayes &
        0.32 \raisebox{0.5ex}{\tiny$^{+0.01}_{-0.01}$} &
        0.30 \raisebox{0.5ex}{\tiny$^{+0.01}_{-0.00}$} &
        0.98 \raisebox{0.5ex}{\tiny$^{+0.02}_{-0.02}$} &
        0.46 \raisebox{0.5ex}{\tiny$^{+0.01}_{-0.01}$} \\

        KNN &
        0.71 \raisebox{0.5ex}{\tiny$^{+0.03}_{-0.02}$} &
        0.54 \raisebox{0.5ex}{\tiny$^{+0.07}_{-0.07}$} &
        0.33 \raisebox{0.5ex}{\tiny$^{+0.05}_{-0.07}$} &
        0.41 \raisebox{0.5ex}{\tiny$^{+0.06}_{-0.06}$} \\

        MLP &
        0.70 \raisebox{0.5ex}{\tiny$^{+0.00}_{-0.02}$} &
        0.50 \raisebox{0.5ex}{\tiny$^{+0.50}_{-0.31}$} &
        0.03 \raisebox{0.5ex}{\tiny$^{+0.03}_{-0.03}$} &
        0.06 \raisebox{0.5ex}{\tiny$^{+0.20}_{-0.06}$} \\

        Quadratic Discriminant &
        0.33 \raisebox{0.5ex}{\tiny$^{+0.01}_{-0.02}$} &
        0.30 \raisebox{0.5ex}{\tiny$^{+0.01}_{-0.01}$} &
        0.97 \raisebox{0.5ex}{\tiny$^{+0.03}_{-0.02}$} &
        0.46 \raisebox{0.5ex}{\tiny$^{+0.01}_{-0.01}$} \\
    \hline
    \end{tabular}
\end{table}

\begin{table}
    \centering
    \caption{Results of the real-life dataset test on the 70-30 distribution trained models to identify SN Ia.}
    \label{ia_7030_val}
    \tiny
    \begin{tabular}{l c c c c}  
    \hline
        Model & Accuracy  & Precision  & Recall & F1\_Score \\
        \hline
        Logistic Regression &
        0.47 \raisebox{0.5ex}{\tiny$^{+0.03}_{-0.02}$} &
        0.25 \raisebox{0.5ex}{\tiny$^{+0.25}_{-0.25}$} &
        0.04 \raisebox{0.5ex}{\tiny$^{+0.04}_{-0.04}$} &
        0.06 \raisebox{0.5ex}{\tiny$^{+0.07}_{-0.06}$} \\

        SVM &
        0.50 \raisebox{0.5ex}{\tiny$^{+0.05}_{-0.01}$} &
        0.40 \raisebox{0.5ex}{\tiny$^{+0.10}_{-0.40}$} &
        0.04 \raisebox{0.5ex}{\tiny$^{+0.96}_{-0.04}$} &
        0.08 \raisebox{0.5ex}{\tiny$^{+0.60}_{-0.08}$} \\

        Decision Trees &
        0.62 \raisebox{0.5ex}{\tiny$^{+0.05}_{-0.06}$} &
        0.67 \raisebox{0.5ex}{\tiny$^{+0.06}_{-0.08}$} &
        0.48 \raisebox{0.5ex}{\tiny$^{+0.10}_{-0.10}$} &
        0.56 \raisebox{0.5ex}{\tiny$^{+0.07}_{-0.09}$} \\

        Random Forest &
        0.64 \raisebox{0.5ex}{\tiny$^{+0.04}_{-0.05}$} &
        0.80 \raisebox{0.5ex}{\tiny$^{+0.05}_{-0.08}$} &
        0.38 \raisebox{0.5ex}{\tiny$^{+0.08}_{-0.10}$} &
        0.51 \raisebox{0.5ex}{\tiny$^{+0.08}_{-0.11}$} \\

        AdaBoost &
        0.62 \raisebox{0.5ex}{\tiny$^{+0.05}_{-0.07}$} &
        0.70 \raisebox{0.5ex}{\tiny$^{+0.07}_{-0.11}$} &
        0.42 \raisebox{0.5ex}{\tiny$^{+0.10}_{-0.18}$} &
        0.53 \raisebox{0.5ex}{\tiny$^{+0.09}_{-0.17}$} \\

        Naive Bayes &
        0.60 \raisebox{0.5ex}{\tiny$^{+0.04}_{-0.02}$} &
        0.56 \raisebox{0.5ex}{\tiny$^{+0.04}_{-0.02}$} &
        0.94 \raisebox{0.5ex}{\tiny$^{+0.04}_{-0.10}$} &
        0.71 \raisebox{0.5ex}{\tiny$^{+0.01}_{-0.02}$} \\

        KNN &
        0.64 \raisebox{0.5ex}{\tiny$^{+0.03}_{-0.06}$} &
        0.73 \raisebox{0.5ex}{\tiny$^{+0.05}_{-0.08}$} &
        0.44 \raisebox{0.5ex}{\tiny$^{+0.06}_{-0.12}$} &
        0.55 \raisebox{0.5ex}{\tiny$^{+0.06}_{-0.12}$} \\

        MLP &
        0.51 \raisebox{0.5ex}{\tiny$^{+0.08}_{-0.02}$} &
        0.50 \raisebox{0.5ex}{\tiny$^{+0.24}_{-0.36}$} &
        0.04 \raisebox{0.5ex}{\tiny$^{+0.24}_{-0.02}$} &
        0.08 \raisebox{0.5ex}{\tiny$^{+0.34}_{-0.04}$} \\

        Quadratic Discriminant &
        0.47 \raisebox{0.5ex}{\tiny$^{+0.14}_{-0.01}$} &
        0.25 \raisebox{0.5ex}{\tiny$^{+0.31}_{-0.14}$} &
        0.04 \raisebox{0.5ex}{\tiny$^{+0.72}_{-0.02}$} &
        0.07 \raisebox{0.5ex}{\tiny$^{+0.58}_{-0.04}$} \\
        \hline
    \end{tabular}
\end{table}

\begin{table}
    \centering
    \caption{Results of the real-life dataset test on the 70-30 distribution trained models to identify SNe Ic-BL.}
    \label{icbl7030_val}
    \tiny
    \begin{tabular}{l c c c c}  
    \hline
        Model & Accuracy  & Precision  & Recall & F1\_Score \\
        \hline
        Logistic Regression & 
        0.72 \raisebox{0.5ex}{\tiny$^{+0.03}_{-0.00}$} & 
1.00 \raisebox{0.5ex}{\tiny$^{+0.00}_{-0.17}$} & 
0.18 \raisebox{0.5ex}{\tiny$^{+0.09}_{-0.00}$} & 
0.31 \raisebox{0.5ex}{\tiny$^{+0.14}_{-0.00}$} \\
SVM & 
0.66 \raisebox{0.5ex}{\tiny$^{+0.35}_{-0.00}$} & 
0.00 \raisebox{0.5ex}{\tiny$^{+0.48}_{-0.00}$} & 
0.00 \raisebox{0.5ex}{\tiny$^{+0.87}_{-0.00}$} & 
0.00 \raisebox{0.5ex}{\tiny$^{+0.00}_{-0.00}$} \\
Decision Trees & 
0.58 \raisebox{0.5ex}{\tiny$^{+0.06}_{-0.06}$} & 
0.38 \raisebox{0.5ex}{\tiny$^{+0.08}_{-0.09}$} & 
0.36 \raisebox{0.5ex}{\tiny$^{+0.09}_{-0.09}$} & 
0.38 \raisebox{0.5ex}{\tiny$^{+0.08}_{-0.09}$} \\
Random Forest & 
0.69 \raisebox{0.5ex}{\tiny$^{+0.02}_{-0.03}$} & 
1.00 \raisebox{0.5ex}{\tiny$^{+0.00}_{-0.00}$} & 
0.14 \raisebox{0.5ex}{\tiny$^{+0.05}_{-0.05}$} & 
0.23 \raisebox{0.5ex}{\tiny$^{+0.07}_{-0.08}$} \\
AdaBoost & 
0.63 \raisebox{0.5ex}{\tiny$^{+0.05}_{-0.06}$} & 
0.42 \raisebox{0.5ex}{\tiny$^{+0.08}_{-0.11}$} & 
0.32 \raisebox{0.5ex}{\tiny$^{+0.09}_{-0.09}$} & 
0.35 \raisebox{0.5ex}{\tiny$^{+0.07}_{-0.09}$} \\
Naive Bayes & 
0.34 \raisebox{0.5ex}{\tiny$^{+0.00}_{-0.03}$} & 
0.34 \raisebox{0.5ex}{\tiny$^{+0.00}_{-0.01}$} & 
1.00 \raisebox{0.5ex}{\tiny$^{+0.00}_{-0.00}$} & 
0.51 \raisebox{0.5ex}{\tiny$^{+0.00}_{-0.01}$} \\
KNN & 
0.65 \raisebox{0.5ex}{\tiny$^{+0.03}_{-0.03}$} & 
0.45 \raisebox{0.5ex}{\tiny$^{+0.09}_{-0.10}$} & 
0.18 \raisebox{0.5ex}{\tiny$^{+0.05}_{-0.09}$} & 
0.28 \raisebox{0.5ex}{\tiny$^{+0.08}_{-0.08}$} \\
MLP & 
0.66 \raisebox{0.5ex}{\tiny$^{+0.02}_{-0.00}$} & 
0.00 \raisebox{0.5ex}{\tiny$^{+0.33}_{-0.00}$} & 
0.00 \raisebox{0.5ex}{\tiny$^{+0.05}_{-0.00}$} & 
0.00 \raisebox{0.5ex}{\tiny$^{+0.00}_{-0.00}$} \\
Quadratic Discriminant & 
0.32 \raisebox{0.5ex}{\tiny$^{+0.02}_{-0.02}$} & 
0.32 \raisebox{0.5ex}{\tiny$^{+0.00}_{-0.01}$} & 
0.91 \raisebox{0.5ex}{\tiny$^{+0.05}_{-0.05}$} & 
0.47 \raisebox{0.5ex}{\tiny$^{+0.01}_{-0.02}$} \\
        \hline
    \end{tabular}
\end{table}
\end{document}